\documentclass[12pt]{article}
\usepackage{amsmath}
\usepackage{graphicx}
\usepackage{natbib}
\usepackage{url} 
\usepackage{booktabs}
\usepackage{latexsym}
\usepackage{amsmath}
\usepackage{ amsthm, amsfonts}
\usepackage{algorithm, algorithmic}  
\usepackage{multicol}
\usepackage{graphicx}
\usepackage{mathtools}
\usepackage{bm, float}
\usepackage{tikz}
\usepackage{tikz-qtree}
\usepackage{etoolbox}
\usepackage{hyperref}
\usepackage{subfigure}
\newcommand{\lGamma}{\text{log}\Gamma}
\newcommand{\mean}[1]{\overline{#1}}

\newcommand{\blind}{0}

\begin{document}

\def\spacingset#1{\renewcommand{\baselinestretch}%
{#1}\small\normalsize} \spacingset{1}

\if0\blind
{
  \title{\bf Tree Estimation and Saddlepoint-Based Diagnostics for the Nested Dirichlet Distribution: Application to Compositional Behavioral Data}
  \author{Jacob Turner \thanks{
    The authors gratefully acknowledge the University PhD Fellowship awarded to Bianca A. Luedeker
by the SMU Moody School of Graduate and Advanced Studies.}\hspace{.2cm}\\
    Department of Mathematics and Statistics, Stephen F. Austin State University\\
    and \\
    Monnie McGee \\
    Department of Statistics and Data Science, Southern Methodist University\\
    and \\
    Bianca Luedeker \\
    Department of Mathematics and Statistics, Northern Arizona University}
  \maketitle
} \fi

\if1\blind
{
  \bigskip
  \bigskip
  \bigskip
  \begin{center}
    {\LARGE\bf Title}
\end{center}
  \medskip
} \fi

\bigskip
\begin{abstract}
The Nested Dirichlet Distribution (NDD) provides a flexible alternative to the Dirichlet distribution for modeling compositional data, relaxing constraints on component variances and correlations through a hierarchical tree structure. While theoretically appealing, the NDD is underused in practice due to two main limitations: the need to predefine the tree structure and the lack of diagnostics for evaluating model fit. This paper addresses both issues. First, we introduce a data-driven, greedy tree-finding algorithm that identifies plausible NDD tree structures from observed data. Second, we propose novel diagnostic tools, including pseudo-residuals based on a saddlepoint approximation to the marginal distributions and a likelihood displacement measure to detect influential observations. These tools provide accurate and computationally tractable assessments of model fit, even when marginal distributions are analytically intractable. We demonstrate our approach through simulation studies and apply it to data from a Morris water maze experiment, where the goal is to detect differences in spatial learning strategies among cognitively impaired and unimpaired mice. Our methods yield interpretable structures and improved model evaluation in a realistic compositional setting. An accompanying R package is provided to support reproducibility and application to new datasets.
\end{abstract}

\noindent%
\spacingset{1.75} 

\section{Introduction}
\label{sec:intro}

Compositional data arise in many applied contexts where the interest lies in the relative contributions of components to a whole. Examples include microbiome analyses \citep{gloor2017}, time-use surveys \citep{vonRosen2023}, comparative glycomics \citep{Bennett2025}, and animal behavioral experiments \citep{Aebischer1993}. Formally, compositional data are defined as a set of random variables where $X_j>0$ and $\sum_{j=1}^p X_j=W$, where $p$ is the number of components and $W$ is a positive real value. The values of $X_j$ can be treated as a discrete count or a continuous measurement.  When measurements are continuous or counts have a large range, the measurements or counts are often divided by $W$, therefore, without loss of generality, $W=1$ and $0 \le X_j \le 1$. 

The Dirichlet distribution (DD) has been used for modeling compositional data due to its convenient closed-form density and natural support on the simplex. The DD has been used to test for the difference in mean vectors for compositional data within a two sample comparison study design \citep{maugard, tml2025}. The most general set of tools for modeling compositional data with DD is Dirichlet regression \citep{hijazi2009modeling}.
 However, the Dirichlet distribution imposes strong constraints on the covariance structure: it assumes that all components are negatively correlated \citep{aitch1985}, and it requires components with the same mean to have identical variances. More specifically, variables with the same mean must have the same beta marginal distributions for the DD \citep{dennis1991}. These limitations can lead to poor model fit in applied settings, where more flexible covariance and variance structures are needed.

Another criticism of the DD is that it does not have the property of subcompositional coherence. Briefly, subcompositional coherence is a property of compositional data such that any statistical inference conducted on a subset of variables renormalized to be its own composition (subcomposition), should be consistent with the inference conducted on the full compositional data set \citep{aitch1982}. Subcompositional coherence is of particular importance when correlations between the variables are of key interest and when not all variables within the composition are readily available to be observed in the sample. 

\citet{aitch1982} advocated statistical analysis of compositional data using log-ratio transformations rather than the use of distributions whose domains are on the simplex such as the DD.  Log-ratio transforms possess the property of subcompositional coherence \citep{aitch1982}.  Transforming compositional data using a log ratio maps the data from a $k-$variate unit simplex and onto $\mathbb{R}^{k-1}$, where multivariate methods based on the Normal distribution can be applied. Therefore, the results can be interpreted on the transformed scale or back-transformed onto the unit simplex using an appropriate parameterization.  When the focus of inference is on changes in the compositional mean between groups or as a function of a covariate, the mean estimates under the transformed model can be mapped back onto the unit simplex through the notion of perturbation.  Perturbation measures changes between the means in a relative fashion rather than raw magnitudes. When compositional data are truly observed from a fixed constraint, such as measuring personal daily activity within a single 24-hour day, raw absolute changes are still meaningful measurements of central tendency.

In addition to Aitchison's work, many authors have proposed richer classes of distributions that contain the Dirichlet distribution as a special case \citep{aitch1985,connor1969,dennis1991,barndorff1991some,ongaro2013generalization}.  While many of these generalizations are not subcompositionally coherent, they can still be helpful when inference is focused on the mean of the composition or when an estimate of the joint and marginal densities is of interest.  One particular generalization is the nested Dirichlet distribution (NDD) \citep{null2009}, which has been utilized in a Bayesian framework for multiple applications, from microbiome studies to sports analytics \citep{gloor2017, tml2025,liu2020empirical,null2009}.  The distribution is defined by using a tree where the choice of tree structure helps dictate the amount of distributional flexibility the random variables possess.  Until recently, there has been little discussion on statistical methodology using  NDD.  The nested Dirichlet distribution has been given multiple names over the years.  It was originally called the Hyper-Dirichlet I distribution \citep{dennis1991} and has also been referred to as the Nested Dirichlet Tree distribution \citep{null2009}. To add to the naming confusion, the NDD also generalizes the Generalized Dirichlet distribution (GDD) \citep{connor1969,dennis1991}. 

\begin{figure}
\centering
\begin{tikzpicture}
\node{}[sibling distance = .7 cm]
    child { node {$X_1$} }
    child { node {$X_2$} }
   child { node {$X_3$} }
   child {node {$X_4$}};
    \node at (0.5,-2,1) {(a)};
\end{tikzpicture}
\qquad
\begin{tikzpicture}
\node{}[sibling distance= 1.2 cm]
    child { node {$N_1$} 
        child { node {$X_1$} }
        child { node {$X_2$} }
        child[missing]
    }
    child { node {$N_2$}
        child[missing]
        child { node {$X_3$} }
        child { node {$X_4$} }
 };
    \node at (0.5,-3.5,1) {(b)};
\end{tikzpicture}
\qquad
\begin{tikzpicture}
\node{}
   child { node {$X_1$} }
    child { node {$N_1$} 
        child { node {$X_2$} }
        child { node {$N_2$} 
        		child{node{$X_3$}}
        		child{node{$X_4$}
            }}};
        \node at (0.5,-5.2,0.5) {(c)};
\end{tikzpicture}
\caption{Comparison of variations of the Nested Dirichlet distribution: (a) standard Dirichlet (b) Nested Dirichlet Distribution (c) Generalized Dirichlet distribution. Each distribution has 5 components. }\label{fig:diffDDs}
\end{figure}
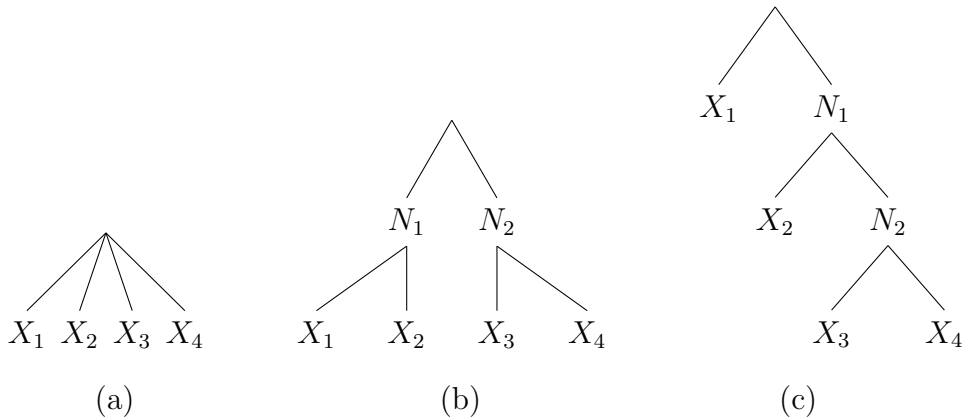

The nested Dirichlet distribution will be formally defined in Section \ref{sec:ndd}. Conceptually, the NDD generalizes the Dirichlet distribution by introducing additional parameters that relax variance and correlation constraints \citep{null2009}. The NDD and its parameters can be visualized and defined using a tree. Figure \ref{fig:diffDDs}  illustrates the distinction between the standard Dirichlet distribution, the nested Dirichlet distribution, and the generalized Dirichlet distribution, where each distribution has $p=5$ variables, also referred to as components: $x_1,x_2,x_3,x_4,x_5$. Figure \ref{fig:diffDDs}(a) represents the standard DD as the tree only contains a root node.  Figure \ref{fig:diffDDs}(b) and (c) contain additional interior nodes labeled $N_1$ and $N_2$; therefore, they represent NDDs with $p=5$ components. In particular, the tree depicted in Figure \ref{fig:diffDDs}(c) contains $p-1$ nodes (including the root) in a cascading effect of binary splits. This characterizes the generalized Dirichlet distribution, which is a subset of the NDD \citep{connor1969}.

\citet{tml2025} proposed using the NDD for two sample comparisons of compositional data and later extended their results to more than two samples \citep{mtl2025}.  The primary contributions of their work include parameter estimation of the NDD, the development of a likelihood ratio test to compare the two mean vectors, and one-at-a-time post hoc confidence intervals for the absolute difference and relative difference in means for each variable. 
When applying their methodology, the authors acknowledge that both maximum likelihood estimation and testing procedures require the analyst to specify the tree in advance.  They recommend that users apply simple distributional rules to suggest and rule out trees by examining sample summary statistics \citep{tml2025}.  Plausible trees can be compared by determining which tree maximizes the log-likelihood.  This approach has a few drawbacks.  The first is that when the number of variables is greater than 5, many plausible trees may exist and the selected tree may not provide the best fit to the data. Secondly, the authors do not propose any sort of residual diagnostics or other numerical metrics, procedures, or visuals to assess the fit of their final model. The authors argued that the NDD was a good fit by simulating random draws using the parameters obtained by the fit and compared summary statistics between the data and random draws.

While, the NDD addresses the lack of flexibility of the DD by introducing a hierarchical tree structure that governs relationships among components, its practical use is limited by the challenges mentioned in \citep{tml2025} and \citep{mtl2025}. First, the NDD requires that a tree structure be specified in advance, which is often infeasible in applied settings where prior knowledge of component relationships is lacking. Second, model diagnostics for the NDD remain underdeveloped. The marginal distributions of the NDD are generally intractable, making it difficult to compute residuals or perform goodness-of-fit assessments. Without such tools, analysts cannot easily evaluate the adequacy of the model or identify poorly fitting observations.

To address these challenges, we introduce a data-driven algorithm for tree structure estimation and propose new diagnostic tools for the NDD. Our approach includes the use of pseudo-residuals derived from a saddlepoint approximation to the marginal distributions. This approximation enables efficient and accurate diagnostics that are computationally feasible and interpretable. In addition, we develop a likelihood displacement measure to detect influential observations. These contributions provide applied researchers with the tools needed to fit and evaluate NDD models in practice.

We demonstrate our methods both with simulated data and with real data from a Morris water maze experiment \citep{maugard}. Morris water maze experiments are designed to assess spatial learning in mice \citep{morris, Morris1984, Tian2019}, and have also been extended to humans using fMRI measurements \citep{Reynolds2019}. For our example, fourteen mice, seven normal mice and seven cognitively impaired mice, were trained to locate a submerged escape platform in a circular pool. More detail is given in Section \ref{sec:wm}, but for now, the amount of time each mouse spent within each of the four sections of a water maze was recorded during testing. Converted to proportions, the data form a four-part composition that reflects each mouse’s learned search strategy. These data motivate the use of the NDD, as they exhibit both compositional constraints and complex correlation structures across quadrants.

In the sections that follow, we present our tree-finding algorithm, propose diagnostics for model evaluation, and demonstrate the practical utility of these methods through simulation and real-data analysis. Code and documentation to reproduce all results are available on GitHub.

\section{The Nested Dirichlet Distribution}\label{sec:ndd}

Before defining the NDD, we first define the Dirichlet density. Let $\mathbf{X}=(X_1,X_2,...,X_p)$, then the density function for a $p-$variate DD is given by
\begin{equation}
f({\bf x}\vert \boldsymbol{\alpha} )=\beta\prod_{j=1}^p x_j^{\alpha_j-1}\qquad 0\le x_j \le 1; j=1, \ldots , p
\label{eq:DD}
\end{equation}
where $\beta=\frac{\Gamma(\Sigma_{j=1}^ {p}\alpha_{j})}{\prod_{j=1}^{ p}\Gamma(\alpha_{j})}$ is the reciprocal of the multivariate beta function and the vector ${\bf x}$ is constrained to sum to one, $\Sigma_{j=1}^p x_j = 1$. For a $p-$variate Dirichlet, the distribution will have $p$ parameters, $\boldsymbol{\alpha}=(\alpha_1,\ldots , \alpha_p )$.  The $\alpha_j$ can be thought of as counts from a prior or current study. It is common to reference $\phi = \sum_{j=1}^{p} \alpha_j$ as the precision parameter because larger values of $\phi$ correspond to smaller variance and less skewness among the variables. The mean for each variable $\pi_j := \frac{\alpha_j}{\phi}$ and the variance is $\sigma_j^2 = \frac{\pi_j(1-\pi_j)}{\phi+1}$; therefore, components with the same mean must also have the same variance.  The covariance between any two components is $\sigma_{jj'}=-\frac{\alpha_j\alpha_{j'}}{\phi(\phi+1)}$ for $j\neq j'$ and is thus non-positive for all cases. The covariance formula also implies that if two pairs of variables in the composition have the same corresponding means, then their correlations must also be the same.

The density function for a $p-$variate NDD, $\mathbf{X}=(X_1,X_2,...,X_p)$, can be defined over any tree with $p$ terminal nodes and $r$ internal nodes. For a given tree, let $\alpha_{k}$ be the parameter associated with the tree edge leading into node $k$, let $T=\{1,2,...,p\}$ be the set of terminal nodes, $I=\{N_1,N_2,...,N_r\}$ the internal nodes, and $R$ the root node.  Additionally, define $C(k)$ to be the immediate children of node $k$ and let $T(k)$ be the set of terminal nodes under node $k$. Lastly, let $\beta(k)=\frac{\Gamma(\Sigma_{j}^ {C(k)}\alpha_{j})}{\prod_j^{ {C(k)}}\Gamma(\alpha_{j})}$ define the reciprocal of the multivariate beta function evaluated using the $\alpha$'s immediately under node $k$. The density function of the NDD is defined as 
\begin{equation}
f({\bf x}\vert \boldsymbol{\alpha})= \beta(R)\left( \prod_{j \in T}x_{j}^{\alpha_{j}-1} \right) \left(\prod_{s \in I} \beta(s) \left(\sum_{j}^{ T(s) }x_{j} \right)^{\Delta(s)} \right) \qquad 0\le x_j \le 1; j=1, \ldots , p
\label{eq:NDD}
\end{equation}
where $\Delta(s)=\alpha_{s}-\Sigma_{k \in C(s)}\alpha_{k}$ are defined over $s \in I$ and $\sum_{j=1}^p x_i=1$. The quantity $\Delta(s)$ is simply the difference between the parameter of an internal node and the sum of its children's parameters.  If $\Delta(s)=0$ for all $s$, then the density reduces to \eqref{eq:DD}. 

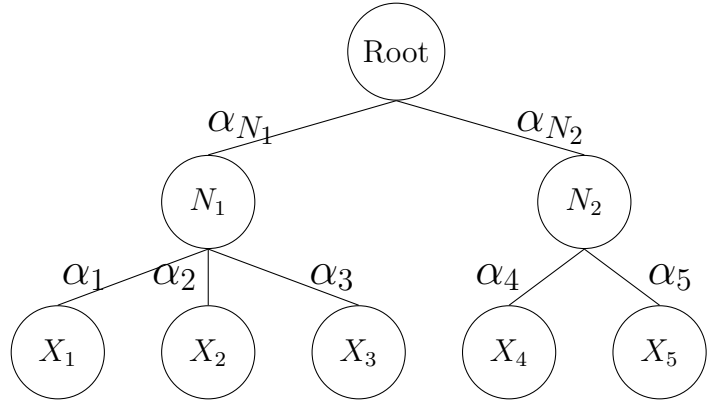
\begin{figure}
    \centering
   \begin{tikzpicture}
    \node[circle, draw, minimum size = 3em]{Root}[sibling distance= 5 cm, level distance = 2 cm]
        child {node[circle, draw, minimum size = 3em] {$N_1$} [sibling distance = 2 cm] 
            child{node[circle, draw, minimum size = 3em] {$X_1$}
            edge from parent node[left, xshift=-0.2cm] {\Large $\alpha_1$}}
            child {node[circle, draw, minimum size = 3em] {$X_2$}
            edge from parent node[left, xshift=-0.0cm] {\Large $\alpha_2$}}
            child {node[circle, draw, minimum size = 3em] {$X_3$}
            edge from parent node[right, xshift=0.2cm] {\Large $\alpha_3$}}
            edge from parent node[left, xshift=-0.2cm]{\Large $\alpha_{N_1}$}
        }
        child {node[circle, draw, minimum size = 3em] {$N_2$} [sibling distance = 2 cm]
            child {node[circle, draw, minimum size = 3em] {$X_4$}
            edge from parent node[left, xshift=-0.2cm] {\Large $\alpha_4$}}
            child{node[circle, draw, minimum size = 3em] {$X_5$}
            edge from parent node[right, xshift=0.2cm] {\Large $\alpha_5$}}
            edge from parent node[right, xshift=0.2cm]{\Large $\alpha_{N_2}$}
        };
    \end{tikzpicture}
    \caption{Tree diagram example for a NDD with corresponding parameters.}
    \label{fig:NDDfigexample}
\end{figure}

An example is provided in Figure \ref{fig:NDDfigexample} where $p=5$ and there are two internal nodes whose immediate parents are the root node.  The NDD density function for the example displayed in Figure \ref{fig:NDDfigexample} is
\begin{equation*}
  f(\bm{x}|\boldsymbol{\alpha}) = \beta^* \prod_{j=1}^p x_j^{\alpha_j-1} \left( \sum_{j=1}^3 x_j \right)^{\alpha_{N_1}-\sum_{j=1}^3 \alpha_j} \left( \sum_{j=4}^5 x_j \right)^{\alpha_{N_2}-\sum_{j=4}^5 \alpha_j} 
\label{eq:nddexample}
\end{equation*}
where $ 0\leq x_j \leq 1\ \text{for}\ j=1, \dots ,p$ and $\boldsymbol{\alpha}$ is the vector of parameters $(\alpha_1,\alpha_2,...,\alpha_5,\alpha_{N_1},\alpha_{N_2})$. The normalizing constant, $\beta^*$, is the product of three Dirichlet normalizing constants:  $\beta(N_1)=\frac{\Gamma(\alpha_1+\alpha_2+\alpha_3)}{\Gamma(\alpha_1)\Gamma(\alpha_2)\Gamma(\alpha_3)}$, $\beta(N_2)=\frac{\Gamma(\alpha_4+\alpha_5)}{\Gamma(\alpha_4)\Gamma(\alpha_5)}$, and $\beta(R)=\frac{\Gamma(\alpha_{N_!}+\alpha_{N_2})}{\Gamma(\alpha_{N_1})\Gamma(\alpha_{N_2})}$.

The fact that the normalizing constant for the NDD distribution consists of a different normalizing constant for each of the interior nodes depicted in Figure \ref{fig:NDDfigexample} offers some insight into the density's original derivation by \citep{dennis1991}. The NDD relies on the assumption of independence of the subcompositions generated from the given tree.  Subcompositions, which we also refer to as branch proportions, are defined for a given nonterminal node $k$ as
\begin{equation}
\mathbf{B}_k=\biggl( \frac{\sum_{j}^{T(A)} X_j}{\sum_{j}^{T(k)} X_j}; A \in C(k)\biggr).
\label{eq:branches}
\end{equation}

The length of the vectors $\mathbf{B}_k$ is equal to the cardinality of $C(k)$ as each individual variable within the vector corresponds to a branch proportion generated by using each of the immediate children of node $k$.  We denote the individual variables within the vector $\mathbf{B}_k$ as $B_{k,A}$ which notationally informs us that the branch proportion corresponds to the edge leading from node $k$ to a child node $A \in C(k)$.

\begin{figure}
    \centering
   \begin{tikzpicture}
    \node[circle, draw, minimum size = 3em]{Root}[sibling distance= 5 cm, level distance = 2 cm]
        child {node[circle, draw, minimum size = 3em] {$N_1$} [sibling distance = 2 cm] 
            child{node[circle, draw, minimum size = 3em] {$X_1$}
            edge from parent node[left, xshift=-0.2cm] {$B_{N_1,1}$}}
            child {node[circle, draw, minimum size = 3em] {$X_2$}
            edge from parent node[left, xshift=-0.0cm] {$B_{N_1,2}$}}
            child {node[circle, draw, minimum size = 3em] {$X_3$}
            edge from parent node[right, xshift=0.2cm] {$B_{N_1,3}$}}
            edge from parent node[left, xshift=-0.2cm]{$B_{R,N_1}$}
        }
        child {node[circle, draw, minimum size = 3em] {$N_2$} [sibling distance = 2 cm]
            child {node[circle, draw, minimum size = 3em] {$X_4$}
            edge from parent node[left, xshift=-0.2cm] {$B_{N_2,4}$}}
            child{node[circle, draw, minimum size = 3em] {$X_5$}
            edge from parent node[right, xshift=0.2cm] {$B_{N_2,5}$}}
            edge from parent node[right, xshift=0.2cm]{$B_{R,N_2}$}
        };
    \end{tikzpicture}
    \caption{Tree diagram example for a NDD with corresponding parameters.}
    \label{fig:NDDbranchexample}
\end{figure}
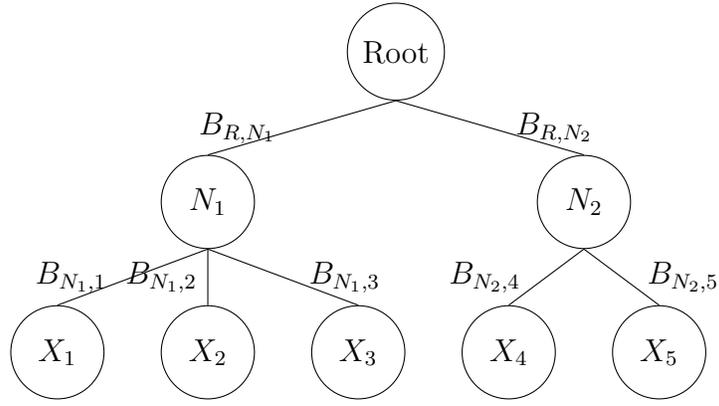

To help visualize the branch transformation, Figure \ref{fig:NDDbranchexample} replaces the parameters associated with each branch in Figure \ref{fig:NDDfigexample} with the branch proportion transformations.  The branch proportions, expressed in terms of the original composition vector $\boldsymbol{X}$ using \ref{eq:branches}, are $\boldsymbol{B}_{N_1}=(\frac{X_1}{X_1+X_2+X_3},\frac{X_2}{X_1+X_2+X_3},\frac{X_3}{X_1+X_2+X_3})$, $\boldsymbol{B}_{N_2}=(\frac{X_4}{X_4+X_5},\frac{X_5}{X_4+X_5})$, and $\boldsymbol{B}_R=(X_1+X_2+X_3,X_4+X_5)$.

In general, the transformation of the original compositional vector $\mathbf{X}$ to the branch proportion vectors, $(\mathbf{B_R},\mathbf{B_{N_1}},...,\mathbf{B_{N_r}})$, is a one-to-one transformation \citep{dennis1991}.  Traditional transformation theory shows that the joint distribution of the branch proportions, derived from \eqref{eq:NDD}, are independent Dirichlet densities,
\begin{equation}
f(\mathbf{b_R},\mathbf{b_{N_1}},...,\mathbf{b_{N_r}} \vert \boldsymbol{\alpha)}=\prod_{k \in \{R \cup I \}} DD(\mathbf{b_k}\vert \boldsymbol{\alpha}_{k})
\label{eq:DirProduct}
\end{equation}
where $\boldsymbol{\alpha}_{k}=\{\alpha_{A}; A \in C(k)\}$ is the set of parameters that correspond to the tree edges leading from node $k$ to its immediate children. Upon reexamining Figure \ref{fig:NDDfigexample} and utilizing Equation \eqref{eq:DirProduct}, the vectors $\boldsymbol{B}_R,\boldsymbol{B}_{N_1},\boldsymbol{B}_{N_1}$ are independently Dirichlet distributed with parameters $\boldsymbol{\alpha}_{N_1}=(\alpha_1,\alpha_2,\alpha_3)$, $\boldsymbol{\alpha}_{N_2}=(\alpha_4,\alpha_5)$, and $\boldsymbol{\alpha}_{R}=(\alpha_{N_1},\alpha_{N_2})$, respectively.

The key result of \eqref{eq:DirProduct} can be realized intuitively using Figure \ref{fig:NDDbranchexample}. Each of the branch proportion vectors,  $\boldsymbol{B}_R,\boldsymbol{B}_{N_1},\boldsymbol{B}_{N_2}$, correspond to a single ``layer" of the tree.  By construction, each branch vector is its own composition and thus, three smaller trees can be observed with no interior nodes, corresponding to the Dirichlet setting depicted in Figure \ref{fig:diffDDs}(a).  This important transformational property provides a straightforward path in estimating the parameters of the NDD and will be discussed in detail in the next section. 

When taking the inverse transformation from branches back to the original space, each $X_j$ is expressed as the product of branch proportions determined by following the branches from the root node to the terminal node corresponding to $X_j$. To express this transformation in general, we must consider an ordered set of nodes, denoted $P_j$ for each $j \in T$, that designates the path from the root node to the immediate parent node of variable $X_j$.  It will also be helpful to define the operation $C_j(k)$  which returns the immediate child of a nonterminal node $k$ in the direct path to terminal node $j$.  

Consider the variable $X_2$ from the NDD model depicted in Figure  \ref{fig:NDDbranchexample} as an example. We have $P_2=\{Root,N_1\}$, $C_2(Root)=N_1$, and $C_2(N_1)=2$.  The variable $X_2$ can then be expressed in terms of branch proportions as 
\begin{align*}
X_2&=B_{R,N_1}B_{N_1,2}\\&=B_{R,C_2(R)}B_{N_1,C_2(N_1)} . 
\end{align*}
For any given tree with terminal nodes $j \in (1,2,...,p)$, the expression for each $X_j$ of the NDD in terms of branch proportions is 
\begin{equation}
   X_j=\prod_{k \in P_j} B_{k,C_j(k)}. 
\label{eq:prodbetas}
\end{equation}

The mean vector $\boldsymbol{\pi}$, variances, and correlations for variables within the NDD model are easily obtained using \eqref{eq:prodbetas} and \eqref{eq:DirProduct}.  The derivations and results, while easy to compute using software, are notationally cumbersome and do not offer much insight into how the specific choices of the tree, and thus $\boldsymbol{\alpha}$, impact the characteristics of the distribution.  For completeness, we provide these details in Supplement S1 along with some general discussions.  At a high level, there are two main points.  In order for variables within a NDD to have the same mean and yet different variances, the two variables must not share the same parent node.  Secondly, for variables to potentially have a positive correlations, the variables must be nested underneath at least one common internal node.   


\section{Parameter Estimation}\label{sec:estimation}

We will now consider parameter estimation under two cases assuming that the data can be modeled using a nested Dirichlet distribution.  The first case is that the tree is known in advance \citep{tml2025}, and the second case is that the tree is unknown. In the second case, the tree must be estimated in addition to the parameters.  First, we review parameter estimation under the standard Dirichlet distribution using maximum likelihood estimators (MLEs).  

The log-likelihood for a set of $p-$variate identically distributed Dirichlet observations can be expressed as
\begin{align}
\begin{split}
    l_D(\bm{\alpha}| x) &= n\lGamma(\sum_{j=1}^{p}\alpha_j) - n\sum_{j=1}^{p} \lGamma(\alpha_{j})\\ &+ n \sum_{j=1}^{p}\alpha_{j}\mean{\text{log}x_j} -n\sum_{j=1}^{p}\mean{\text{log}x_j}
    \label{eq:DDloglike}
\end{split}
\end{align}
where $x$ is an $n\times p$ matrix of sample data and the quantity $\mean{\text{log}x_j}=\frac{1}{n}\Sigma_{i=1}^n log x_{ij}$ for $j=1,...,p$.

The MLE, $\bm{\hat{\alpha}}$, is found by maximizing equation \ref{eq:DDloglike} numerically via Newton's method or using an intercept only Dirichlet regression model \citep{minka2000, narayanan, ronning,hijazi2009modeling}. It should be noted that the last term in the likelihood is constant with respect to $\bm{\alpha}$ and thus plays no role in the maximization \citep{minka1999,hijazi2009modeling}.  We will denote $l_D^{*}(\bm{\alpha}| x)$ as the adjusted log-likelihood for the Dirichlet which simply removes the $-n\sum_{j=1}^{p}\mean{\text{log}x_j}$ term from \eqref{eq:DDloglike}.

\subsection{Parameter estimation of the NDD (tree is known)}
  MLEs for the NDD cannot be obtained unless a tree, $\tau$, is specified.  We emphasize this point by denoting the log-likelihood function for the NDD as $l_{ND}(\bm{\alpha}|x,\tau)$. With $\tau$ given, the sets of interior nodes, $I$, along with the functions $C(k),T(k),\beta(k),\Delta(k)$, detailed in Section \ref{sec:ndd}, are fully defined. When $\tau$ is specified with no interior nodes, as in Figure \ref{fig:diffDDs}(a), $l_{ND}(\bm{\alpha}|x,\tau)$ reduces to $\eqref{eq:DDloglike}$. Using \eqref{eq:NDD}, the log-likelihood can be expressed as
\begin{equation}
    l_{ND}(\bm{\alpha}\vert x,\tau) = \sum_{s \in \{R \cup I \}} \log{\beta(s)} + n\sum_{j \in T} \alpha_{j}\mean{\text{log}x_j}+n\sum_{s \in I} \Delta(s)\mean{\text{log}x_{T(s)}}-n\sum_{j\in T}\mean{\text{log}x_j}
\label{NDDloglike}
\end{equation}
where the quantity $\mean{\text{log}x_{T(s)}}=\frac{1}{n}\sum_{i=1}^n log(\sum_{j \in T(s)} x_{ij})$.  Rather than trying to maximize $l_{ND}$ directly, we can rewrite \eqref{NDDloglike} in terms of Dirichlet adjusted log-likelihoods.  To achieve this, we must express the observed data values in terms of their branch proportions using \eqref{eq:prodbetas}.  Substituting \eqref{eq:branches} into \eqref{NDDloglike} and with some careful algebra, the likelihood reduces to
\begin{equation}
    l_{ND}(\bm{\alpha}\vert x,\tau) = \sum_{k \in \{R \cup I \}} l_D^*(\bm{\alpha}_k|b_k)-n\sum_{j\in T}\mean{\text{log}x_j}
\label{NDDloglike2}
\end{equation}
where $b_k$ is the observed matrix of branch proportions generated from node $k$, derived from the original compositional data matrix $x$. Since $l_{ND}$ is equivalent to the sum of adjusted Dirichlet log-likelihoods, maximizing over $\bm{\alpha}$ can be completed by maximizing subsets of the parameters using standard estimation techniques for the Dirichlet.  The simplicity of Equation \ref{NDDloglike2} provides a straightforward procedure in computing the MLEs of a NDD for any specified tree $\tau$.

\subsection{When the tree is unknown}
In practice, the tree $\tau$ might not be known in advance or easily determined from subject matter knowledge or characteristics of the observed data.  In this setting, the tree itself must be estimated in addition to the parameters. One approach is to consider all possible trees and compute the log-likelihood for each tree. The selected tree is the one that maximizes the log-likelihood or an information criterion such as AIC or BIC that penalizes for adding too many interior nodes \citep{tml2025}. For a 4-variate tree ($p = 4$), there are 29 trees to search; however, the number of trees increases combinatorially with the number of variables. A 5-variate NDD has well over 200 possible trees to consider.  If we only consider the cascading trees obtained via the generalized Dirichlet distribution, the number of possible trees is $\frac{p!}{2}$, where $p$ is the number of variables.  While the restriction to cascading trees may alleviate some computational time, it could also lead to considering a model that cannot capture the specific mean, variance, or correlation behavior that is exhibited by the data being fitted.

Another approach for reducing the number of possible trees to be considered is to use sample statistics from the data \citep{tml2025}. As noted in Section \ref{sec:ndd} and in Supplement S1, if two variables have consistent sample means but drastically different variances, the two variables should not share the same parent node.  For variables that exhibit positive sample correlations, the variables should be nested underneath at least one common internal node.   The remaining ``plausible" trees that capture the observed characteristics of the data are fitted via maximum likelihood, and the tree with the largest likelihood is chosen as the estimate for the true tree $\tau$.  Narrowing the possible trees via sample statistics is a somewhat {\it ad hoc} approach; therefore, it is possible to miss a candidate tree that provides a better fit to the data. Clearly, an algorithmic data-driven approach to estimating $\tau$ would ameliorate some of the {\it ad hoc} nature of previous approaches. 

An algorithm to determine goodness of fit using the generalized Dirichlet distribution was previously proposed for the analysis of glycan chromatography data \cite{galligan2013greedy}.  In \cite{minka1999}, a greedy tree-searching algorithm was proposed in the Bayesian context when observing multinomial count data.  Our approach for estimating the tree will follow the same core principles as \cite{minka1999} discusses. We will assume that the data follow a NDD and provide the necessary technical details to implement the approach in practice.  

\begin{figure}
\centering
\includegraphics[scale=.34]{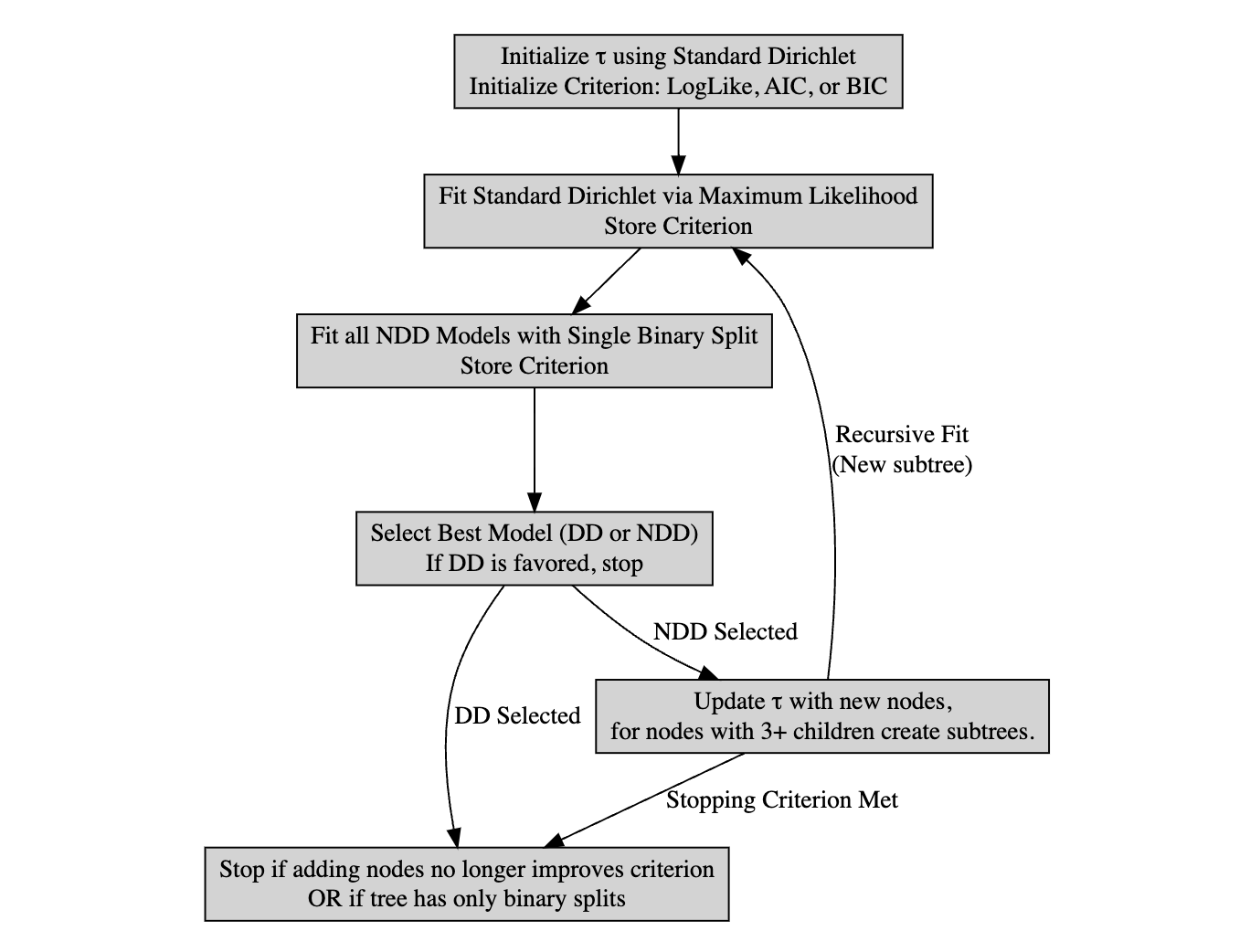}
\caption{Flow chart of the tree-finding algorithm. The algorithm will fit either a DD or NDD to data using maximum likelihood, AIC or BIC as the criterion.}\label{fig:treeAlg}
\end{figure}

While the NDD generalizes the standard Dirichlet, the NDD can also generalize other NDD distributions \cite{minka1999}.  More specifically, the NDD for any given tree can be expressed by a larger tree in which additional interior nodes are added so that the tree is compromised entirely of binary splits.  
By working with binary splits within $\tau$, we can use a greedy, top-down approach for searching candidate trees to find the best fit to the data. This approach is similar to the implementation of regression and classification trees for modeling a response. The algorithm is outlined in Figure \ref{fig:treeAlg}. As an initialization step, we specify $\tau$ by fitting a DD to the data and we use the initial DD to calculate a log-likelihood or other information criterion.  We then fit all possible NDD distributions corresponding to a single binary split at the root node and select the best model using the fit criterion.  If the DD distribution is favored, the algorithm stops.  If one of the nested Dirichlet distributions is favored, $\tau$ is updated to include the interior node(s) and the data are converted to branch proportions corresponding to the optimal binary split. For any branch proportions vectors with more than two variables, we treat them as a new compositional data set and repeat the process, fitting them with a DD and single-binary split NDD's to determine if additional nodes are required. The algorithm stops when either no additional interior nodes are deemed necessary or the tree contains only binary splits.

\begin{figure}
\centering
\begin{tikzpicture}
\node{}[sibling distance = .7 cm]
    child { node {$X_1$} }
    child { node {$X_2$} }
   child { node {$X_3$} }
   child {node {$X_4$}}
   child {node {$X_5$}};
    \node at (0.5,-2,1) {(a) Initial $\tau$};
\end{tikzpicture}
\qquad
\begin{tikzpicture}
\node{}[sibling distance= 1.2 cm]
    child { node {$N_1$} 
        child { node {$X_1$} }
        child { node {$X_3$} }
        child { node {$X_4$}}
        child[missing]
    }
    child { node {$N_2$}
        child[missing]
        child { node {$X_2$} }
        child { node {$X_5$} }
 };
    \node at (0.5,-3.5,1) {(b) First update to $\tau$};
\end{tikzpicture}
\qquad
\begin{tikzpicture}
\node{}[sibling distance= 1.2 cm]
    child { node {$N_1$} 
        child { node {$N_3$}
            child { node {$X_1$} }
            child { node {$X_3$} }
        }
        child { node {$X_4$}}
        child[missing]
    }
    child { node {$N_2$}
        child[missing]
        child { node {$X_2$} }
        child { node {$X_5$} }
 };
    \node at (0.5,-5,1) {(c) Second update to $\tau$};
\end{tikzpicture}
\caption{Hypothetical example of a top-down, greedy search of tree candidates. In step 1 an initial fit using the standard Dirichlet is considered.  Step 2 depicts a better fitting NDD using a single binary split at the root node.  Step 3 applies the splitting algorithm on the branch proportions generated from $X_1$,$X_3$, and $X_4$ in which an additional internal node $N_3$ was favored.
\label{fig:AlgSteps}}
\end{figure}
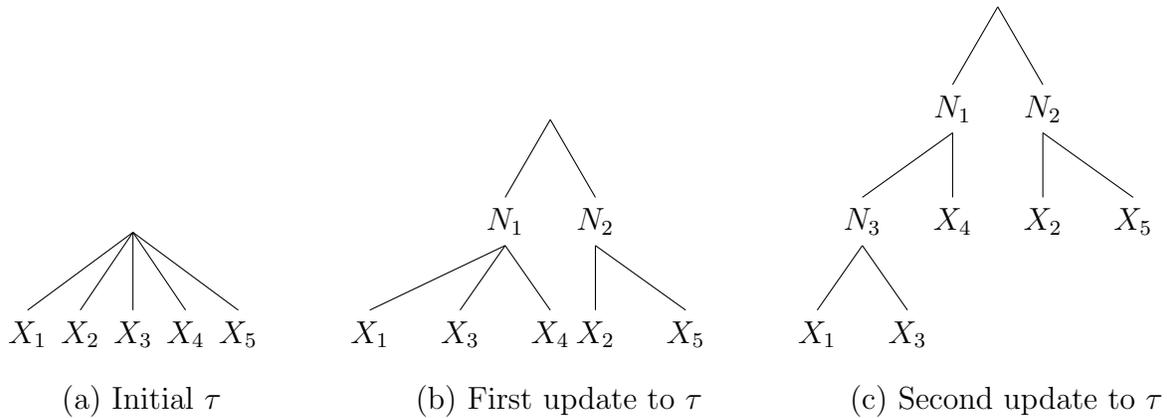

Figure \ref{fig:AlgSteps} illustrates an example of the first three iterations of the fitting process for a 5-variate compositional data set. Figure \ref{fig:AlgSteps}(a) depicts the initialization step of the algorithm in which the standard Dirichlet model is assumed and a fit criterion such as the log-likelihood, AIC, or BIC is stored.  With the simplest tree fitted to the five variables as a baseline model, all possible NDD trees involving a single binary split at the root node are fitted with their corresponding criterion. In our example, it was determined that the maximum log-likelihood value (or smallest AIC/BIC) occurred by incorporating two internal nodes in which $X_1$, $X_3$, and $X_4$ are the children of $N_1$ and the remaining variables are children of $N_2$.  Since there are only 2 variables under $N_2$, no additional nodes can be added.  However, for node $N_1$, we can convert the 3 variables to branch proportion data ,$b_{N_1}$, and repeat the fitting process.  If a standard Dirichlet is favored over any NDD fit using the new subcomposition, then the algorithm will stop, and the tree in Figure \ref{fig:AlgSteps}(b) will be selected.  However, if it is determined that a NDD was a better fit to the subcomposition where $X_1$ and $X_3$ were children of a single interior node, the inclusion of another interior node requires an update of $\tau$ and is depicted in Figure \ref{fig:AlgSteps}(c).  At this point, the algorithm stops since the tree is entirely comprised of binary splits; therefore, the algorithm does not allow inclusion of any additional nodes.   

Although the tree-finding algorithm is relatively straightforward in principle, it is necessary to highlight a few technical points to help ensure reproducibility and maintain algorithmic efficiency. When fitting NDD models to determine whether an additional interior node should be included, there are two general tree structures that yield a single binary split at the root node. The first structure occurs when there are two interior nodes as depicted in Figure \ref{fig:AlgSteps}(b). For a $p$-variate composition, there are $\binom{p}{2}$ unique trees when $p$ is odd and $0.5\binom{p}{2}$ when $p$ is even.  The second structure to consider is when a single variable is a child of the root node and the remaining variables are nested under a single interior node, in which case there are $p$ unique trees to consider.  The number of total trees for calculation and comparison of fit metrics is largest at the initial step of the fitting process and is thus more computationally expensive. As the algorithm recursively considers adding additional interior nodes by investigating the subcompositions, the dimension of the subcompositions reduces to $\lceil\frac{p}{2} \rceil$, or, at worst, to $p-1$.  As additional nodes are added further down the tree, the dimension of the branch proportion vectors decreases. 

It is also important to keep in mind the relationship between the log-likelihood for the NDD distribution, $l_{ND}(\bm{\alpha}\vert x,\tau)$, and the adjusted log-likelihood of the Dirichlet under the branch proportion transformation as expressed in \eqref{eq:DDloglike}.  The fit metrics to compare and decide whether a binary split should occur must depend on comparing $l_{ND}(\bm{\alpha}\vert x,\tau)$ with the different possible $\tau$'s while holding the data $x$ fixed.  Because of the independence of the branch proportions and the fact that they follow the Dirichlet distributions, it can be tempting to obtain the log-likelihood for each given Dirichlet fit of the branches and add the log-likelihoods for each node.  However, with this na\"{i}ve approach, the additional constant removed from the adjusted DD log-likelihood, $l^*_D(\mathbf{\alpha}|x)$, remains. The sum of these constant terms across multiple nodes is typically much smaller than the true NDD log-likelihood constant, $-n\sum_{j \in T} \mean{\text{log}x_j} $. as a result, the na\"{i}ve approach often favors the Dirichlet over the NDD model even when it is clear from either data visualization or summary statistics that a standard Dirichlet fit would not be appropriate.  In summary, comparing models by adding log-likehoods of each branch proportion DD fit is not equivalent to the log-likelihood of the NDD.  It is thus not a recommended strategy to assess the overall fit of competing NDD models.

Equation \eqref{NDDloglike2} also highlights the important fact that a recursive, top-down tree search is viable and that we do not need to monitor the full log-likelihood $l_{ND}(\bm{\alpha}\vert x,\tau)$ in order to maximize it which is implied by the algorithm. To demonstrate this, consider the tree fit examples pictured in Figure \ref{fig:AlgSteps}.  The log-likelihood for the tree in Figure \ref{fig:AlgSteps}(b), denoted $l_{ND}(\mathbf{\alpha}_{(b)}|x,\tau_{(b)})$ is 
\begin{align}
    \begin{split}
    l_{ND}(\mathbf{\alpha}_{(b)}|x,\tau_{(b)})=&\,l^*_D(\alpha_{R(b)}|b_{R(b)})+l^*_D(\alpha_{N_1(b)}|b_{N_1(b)})+l^*_D(\alpha_{N_2(b)}|b_{N_2(b)})\\ & +n\sum_{j=1}^5 \mean{\text{log}x_j}.
    \end{split}
\end{align}
Similarly, the log-likelihood for the tree in Figure \ref{fig:AlgSteps}(c) is 
\begin{align}
    \begin{split}
        l_{ND}(\mathbf{\alpha}_{(c)}|x,\tau_{(c)})=&\,l^*_D(\alpha_{R(c)}|b_{R(c)})+l^*_D(\alpha_{N_1(c)}|b_{N_1(c)})+l^*_D(\alpha_{N_2(c)}|b_{N_2(c)}) \\
        & +l^*_D(\alpha_{N_3(c)}|b_{N_3(c)})+n\sum_{j=1}^5 \mean{\text{log}x_j}.
    \end{split}
\end{align}   
Upon examination, the likelihoods share the same components for the root node, $N_2$, and the constant term.  Because the parameters of any NDD distribution are estimated independently for the branch proportions, obtaining the MLEs for the tree in Figure \ref{fig:AlgSteps}(c) requires estimating only the parameters $\alpha_{N_1(c)}$ and $\alpha_{N_3(c)}$ as the remaining parameters are already obtained from a previous fit using the tree in Figure \ref{fig:AlgSteps}(b).  Because the MLEs of the Dirichlet must be computed numerically, reducing any redundant calculations is imperative.

Lastly, each interior node that is added to a NDD fit corresponds to one additional parameter to be estimated.  We will always be able to increase the log-likelihood of the NDD by adding more internal nodes that are not necessary. To remedy this, for any given set of trees, we can easily fit each one and evaluate AIC or BIC metrics that penalize the log-likelihood based on the number of parameters included.  It is more technically challenging and computationally intensive to evaluate AIC for the different tree fits due to the recursive nature of the algorithm.  For the sake of computational efficiency, we implement AIC and BIC calculations as part of the recursive process.  Consequently, the penalization obtained via AIC is not in the global sense of the entire fit, but rather a local fit of the composition currently being evaluated for internal node inclusion.  Anecdotally, we have found this to work relatively well with synthetic data.  As expected, using the BIC metric will typically produce tree fits with fewer nodes than will using AIC or an non-penalized log-likelihood.


\section{Assessing the fit}
An important task when fitting a distribution to a sample of data is to assess the goodness of fit.  Fit assessment is typically done by using either a hypothesis test or some form of visual diagnostics, such as quantile-quantile plots or residual diagnostic plots, as in multiple linear regression or generalized linear models.  For assessing the model fit for a NDD, we propose four techniques. The first two approaches allow for visual assessment of the marginal fits, while the remaining two allow for assessment of potential outliers in a joint fashion.

\subsection{Assessing the Fits Marginally}\label{sec:diag}
To motivate this section, we will introduce the notion of a pseudo-residual computed for the DD model, \cite{hijazi2006residuals}, and then extend this notion to the NDD case. For the DD model with precision parameter $\phi$, the marginal density for each $X_j$, $f_{X_j}(x_j|\bm{\alpha})$, is distributed beta with shape parameters $a=\alpha_j$ and $b= \phi - \alpha_j$.  If the DD is the true model, then a good approximation to each marginal density can be obtained by substituting the MLEs into the marginal density function $\hat{f}_{X_j}=f_{X_j}(x_j|\bm{\hat{\alpha}})$.  When the sample size is large enough, it can be helpful to provide histograms of each variable and overlay the fitted density curves to assess the fit. Alternatively, one could generate realizations of equivalent sample sizes from the fitted model, examine the realizations both visually and numerically.  The characteristics obtained in the realizations can then be compared to determine whether the key characteristics of the marginal distributions and the correlation behavior between components are present \citep{tml2025}.

As noted in \cite{hijazi2006residuals}, a pseudo-residual for the $x_{ij}$ observation in the observed data matrix $x$ can be obtained via the probability integral transform by setting $r_{ij}=\Phi^{-1}(\hat{F}_{X_j}(x_{ij})),$ where $\hat{F}_{X_j}$ is the estimated distribution function for $X_j$ and $\Phi^{-1}(x)$ is the inverse distribution function of the standard normal.  If the model is correctly specified, then the pseudo-residuals corresponding to each variable $X_j$ should follow a standard normal distribution and can be investigated with a Q-Q plot.  Any gross departures from the 45 degree line can indicate a poor fit or help identify outliers in the univariate sense.

The pseudo residual technique can be applied also to the NDD case with one major hurdle. Recall from \eqref{eq:prodbetas} that, for the NDD, each variable is expressed as either a beta distribution if $X_j$ has the root node as its immediate parent or, more generally, as the product of independent beta distributions. The latter does not have a tractable distribution function.  Because of this intractability, there has been very little discussion on the marginal distributions of the NDD except for the special case when the parameters are integers \cite{dennis1991}.  However, the distribution of the product of independent beta distributions plays a key role in other settings such as computing p-values when performing the Wilks' likelihood ratio test in MANOVA and its distribution function can be approximated quite accurately using saddlepoint approximations \citep{butler}.  In Supplement S2, we show how saddlepoint approximations can be employed to obtain highly accurate approximations for the marginal density and distribution functions of the NDD.  The saddlepoint approximation to the distribution function from the NDD fit can then be substituted in the place of $\hat{F}_{X_j}$ when computing $r_{ij}=\Phi^{-1}(\hat{F}_{X_j}(x_{ij}))$ to obtain pseudo-residuals.       

\subsection{Detecting Outliers Jointly}
Detection of outliers in multivariate data, particularly those that are outliers in multiple dimensions, is a difficult task. In multivariate data, outliers can be extreme in multiple directions and in increasing numbers of directions \citep{rocke1996}. Assessment of whether an observation drastically influences the overall fit (a multivariable leverage point) is also needed.  To meet this need, we propose the use of a leave-one-out likelihood distance metric proposed for Dirichlet regression \citep{hijazi2006residuals}, which has also been called likelihood displacement: \citep{cook1988}.
\begin{equation}
\label{eqn:likedist}
LD_i=2 \bigg( l_{ND}(\hat{\bm{\alpha}}\vert x,\tau)-l_{ND}(\hat{\bm{\alpha}}_{(i)}\vert x,\tau) \bigg), 
\end{equation}
\noindent where $\hat{\bm{\alpha}}_{(i)}$ is the MLE for the NDD with the $i^{th}$ observation removed.  While this approach can be computationally intensive for large data sets, the information it provides is valuable for ascertaining the influence of each observation. To reduce computational time for larger data sets, parallel computing can be utilized, although we have not directly implemented this strategy. 

For larger data sets, we can focus on the identification of outliers through computing Aitchison's distance, defined as 
\begin{equation}
\label{eqn:aitchdist}
D_i=\Delta(\bm{x_i},\bm{\hat{\pi}})=\bigg[ \sum_{j=1}^p \Big(\log \frac{x_{ij}}{g(\bm{x_i})}-\log \frac{\bm{\hat{\pi}_j}}{g(\bm{\hat{\pi}}))} \Big)^2   \bigg]^{1/2}
\end{equation}
\noindent where $g()$ is the geometric mean. This distance is computed between the mean estimate from the fit and each observation, $\bm{x_i}$ \citep{aitch1985}.

\section{Analysis Examples}
In this section, we demonstrate the general utility of the tree finding algorithm using our proposed diagnostics as well as traditional summary statistics. We will begin by using a large synthetic data set to provide clear discussions and illustrations in a controlled setting.  We follow up by reanalyzing data from a water maze experiment conducted on 14 mice under two conditions: normal cognitive function and cognitively impaired \cite{maugard, tml2025}. We further show the utility of diagnostic graphics and tests on the water maze data by inserting an outlier in the data. Our diagnostics are able to find the outlier. Because the number of tables and figures from our tree-finding algorithm and diagnostics tests is large, we have placed an exhaustive investigation of each data set in an R markdown file available in the Supplementary Material. The rest of this section shows excerpts from the full investigation.

\subsection{Synthetic Data}
For our first example, consider the tree in Figure \ref{fig:NDDfigexample} with components $(X_1, X_2, X_3, X_4, X_5)$ and parameter vector $\boldsymbol{\alpha} =(\alpha_1,\alpha_2,\alpha_3,\alpha_4,\alpha_5,\alpha_{N_1},\alpha_{N_2})$ with values $(0.5,1.5,2,10,10,8,2)$, respectively. The parameter values were selected so that the corresponding mean is $\boldsymbol{\pi}=(.1,.3,.4,.1,.1)$ and $SD(\boldsymbol{X})=(.121,.181,.191,.065,.065)$.  

According to Figure \ref{fig:NDDfigexample}, components $X_4$ and $X_5$ are nested under $N_2$ and have the same mean of 0.1; therefore, their distributions are identical. $X_1$ also has a mean of 0.1, but is nested under $N_1$. Therefore, $X_1$ is governed by a different set of parameter values and its standard deviation is approximately double that of $X_4$ (and $X_5$).  Additionally, $\Delta(N_2)=2-(10+10)=-18$ which helps to generate a strong positive correlation of 0.697 between $X_4$ and $X_5$.  Similarly, $\Delta(N_1)=6$, which generates stronger negative correlations between $X_1$, $X_2$, and $X_3$ than is expected be observed for a DD. 

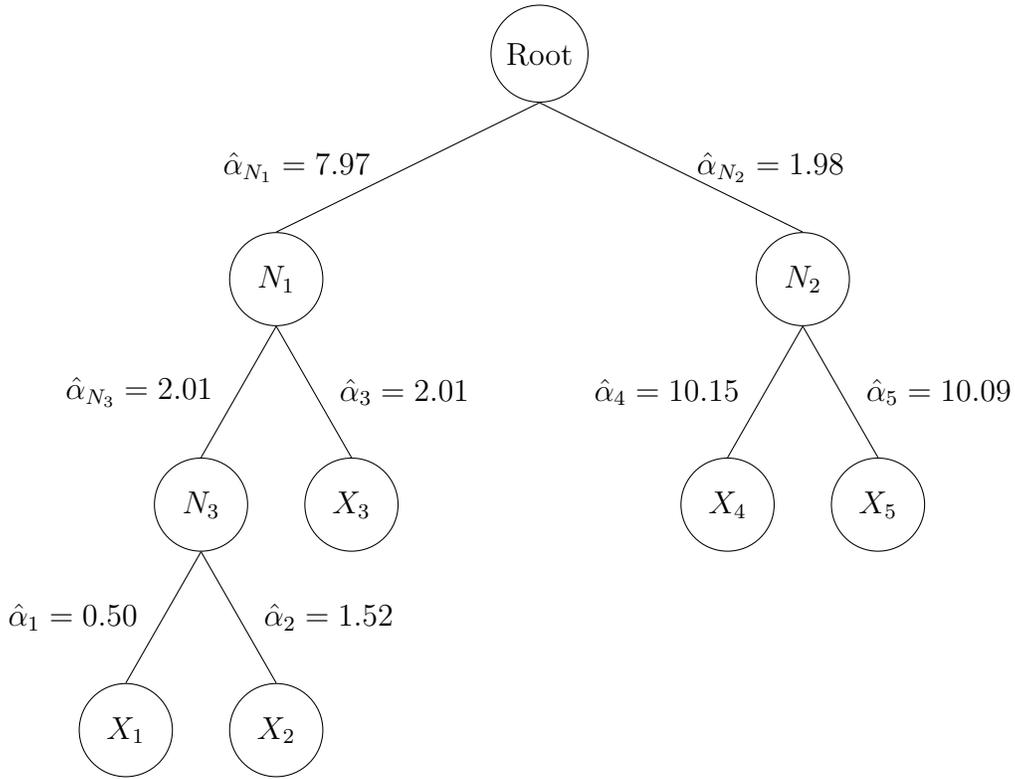
\begin{figure}
    \centering
   \begin{tikzpicture}
    \node[circle, draw, minimum size = 3em]{Root}[sibling distance= 7 cm, level distance = 3 cm]
        child {node[circle, draw, minimum size = 3em] {$N_1$} [sibling distance = 2 cm] 
            child{node[circle, draw, minimum size = 3em] {$N_3$}
                [sibling distance = 2 cm]
                    child{node[circle,draw,minimum size= 3em]{$X_1$}
                    edge from parent node[left,xshift=-.2cm]{ $\hat{\alpha}_1=0.50$}}
                    child{node[circle,draw,minimum size= 3em]{$X_2$}
                    edge from parent node[left,xshift=2.2cm]{ $\hat{\alpha}_2=1.52$}}
            edge from parent node[left, xshift=-0.2cm] { $\hat{\alpha}_{N_3}=2.01$}}
            child {node[circle, draw, minimum size = 3em] {$X_3$}
            edge from parent node[right, xshift=0.2cm] { $\hat{\alpha}_3=2.01$}}
            edge from parent node[right, xshift=-2.6cm]{ $\hat{\alpha}_{N_1}=7.97$}
        }
        child {node[circle, draw, minimum size = 3em] {$N_2$} [sibling distance = 2 cm]
            child {node[circle, draw, minimum size = 3em] {$X_4$}
            edge from parent node[left, xshift=-0.2cm] {$\hat{\alpha}_4=10.15$}}
            child{node[circle, draw, minimum size = 3em] {$X_5$}
            edge from parent node[right, xshift=0.2cm] { $\hat{\alpha}_5=10.09$}}
            edge from parent node[right, xshift=0.2cm]{ $\hat{\alpha}_{N_2}=1.98$}
        };
    \end{tikzpicture}
    \caption{Estimated tree and parameter estimates of the NDD fit to the synthetic data set.}
    \label{fig:SyntheticFit}
\end{figure}

To demonstrate the utility of the methods, we generated 1000 random draws from the NDD specified in Figure \ref{fig:NDDfigexample} with parameter values given in the first paragraph of this section.  We then applied the tree-finding algorithm to estimate both the tree and its parameters.  The summary of the fit is displayed in Figure \ref{fig:SyntheticFit} with the estimated parameters listed along the edges. 

Overall the tree finding algorithm includes interior nodes in the appropriate places given the true tree. However, the algorithm overparameterizes the true distribution by adding an unnecessary interior node, $N_3$.  Note that the estimate of $\Delta(N_3)$ for node $N_3$ is $\hat{\alpha}_{N3}-\hat{\alpha}_{1}-\hat{\alpha}_{2}= -0.0234$. Because $\Delta(N_3)$ is close to 0, the tree in Figure \ref{fig:SyntheticFit} offers no additional flexibility compared to the tree with $N_3$ removed.  Therefore, the tree finding algorithm suggests a tree that generalizes the true tree, and the parameter estimates suggest that the inclusion of node $N_3$ does not contribute to the overall fit. Therefore, the tree-finding algorithm, while not finding the exact tree, finds one that is close, and often generalizes the true tree.

Table \ref{tab:Synthetic} provides a comparison of the mean and standard deviation estimates for all five components for both distributional fits, along with the true parameter values. Estimates of the means and standard deviations from the NDD fit in Figure \ref{fig:NDDfigexample} are consistent with the true values in the first column of Table \ref{tab:Synthetic}. The correlations among the components are also consistent made evident by the relative closeness of the $\bf \hat{\alpha}$'s displayed in Figure \ref{fig:SyntheticFit} to their true values. 
\begin{table}[!htb]
\centering
\begin{tabular}{rrrrr}
  \hline
 & Truth & NDD & DD \\ 
  \hline
  $X_1$ & .1(.12) & .099(.12) & .079(.09) \\ 
  $X_2$ & .3(.18) & .301(.18) & 0.28(.15)  \\ 
  $X_3$ & .4(.19) & .401(.19) & .383(.16)  \\ 
  $X_4$ & .1(.07) & .100(.07) & .129(.11) \\ 
  $X_5$ & .1(.07) & .099(.07) & .128(.11) \\ 
   \hline
\end{tabular}\label{tab:Synthetic}
\caption{Comparison of mean ( standard deviation) from the true fit with estimates from the NDD and DD fits.  The NDD fit generally gives more accurate and precise estimates than the DD fit.}
\end{table}
\begin{figure}
    \begin{centering}
    \includegraphics[scale=.10]{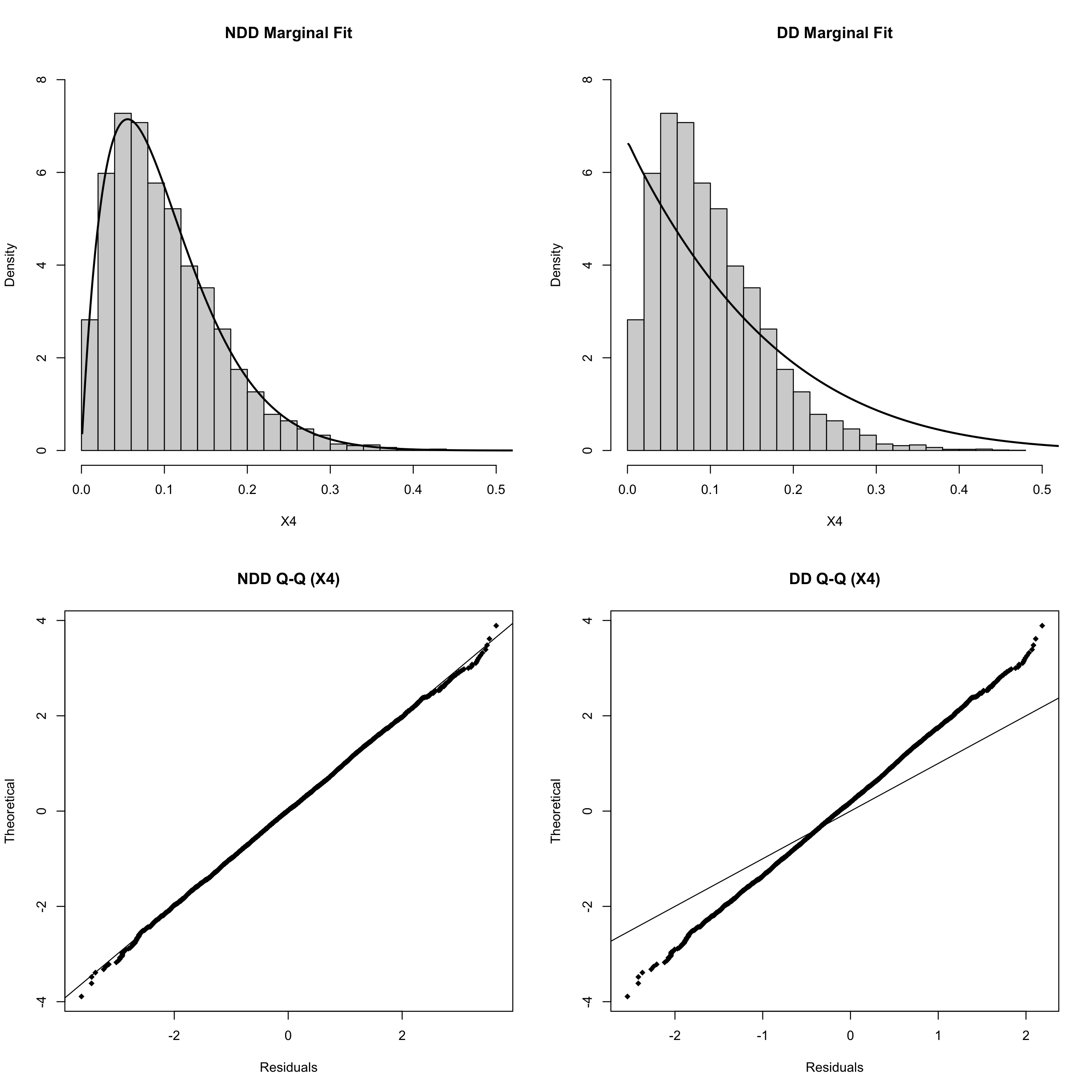}
    \caption{Residual and marginal estimates from NDD and DD fits using the synthetic data.}\label{fig:syntheticcompare}
    \end{centering}
\end{figure}

To demonstrate the utility of both the residual and marginal derivations, we fit a DD model to the synthetic data set and compared the pseudo-residuals and marginal fits from the NDD model fit proposed by the algorithm.  Figure \ref{fig:syntheticcompare} provides this comparison, specifically for $X_4$. It is clear in both the marginal densities, as well as the pseudo-residual plots, that the NDD model is a more adequate fit. Because $X_1$, $X_4$ and $X_5$ have similar sample means but quite different sample standard deviations and marginal shapes, the DD fit cannot take this into account. As depicted in Table \ref{tab:Synthetic}, the DD model drastically overestimates the means of $X_4$ and $X_5$, and underestimates the mean of $X_1$.  

\subsection{Water Maze}\label{sec:wm}

To illustrate the tree-finding algorithm on a real data set, we turn to an analysis of data from an experiment involving a water maze \citep{maugard}. These data were previously used to illustrate the two-sample likelihood ratio test for compositional data based on the NDD \citep{tml2025}. The water maze data had been analyzed with a DD, but there is a positive correlation between two of the components; therefore, an NDD is more suitable. However, the water maze data does not have a pre-defined tree. Previous work used a brute-force method of creating all possible trees and finding the likelihood for each one. The tree with the maximum likelihood was chosen. This method was possible because the water maze data have only 4 components. Here, we use the tree finding algorithm to obtain a tree with log-likelihood as the metric for adding internal nodes. The algorithm produces the same tree as was found by the brute-force method, and we can use diagnostics developed in Section \ref{sec:diag} to show that this tree is a good fit for the data. 

Water maze experiments have been used in a variety of disciplines to examine the ability of mice to use and remember distal cues (cues outside the maze) to perform various tasks \citep{morris, Morris1984, Tian2019, Reynolds2019}. A water maze consists of a circular tank with a platform in one quadrant of the tank. If the mouse finds the platform, it can escape the tank. The quadrant containing the platform is labeled the target quadrant (TQ). Quadrants adjacent to TQ are labeled AQ1 and AQ2, and the quadrant opposite the platform is labeled opposite quadrant (OQ). The tank is not physically divided into quadrants. Quadrants are references for the position of the mouse within the tank. Each mouse is placed in OQ and allowed to swim for a given amount of time or until it locates the platform. Later, the mouse is reintroduced into the maze and is allowed to swim until finding the platform. 

For this specific experiment, researchers recorded the amount of time spent in a Morris water maze for a total of 14 mice. Seven of the mice were normal mice (WT), and others had brain impairment similar to what would be seen in Altzheimer's disease (Tg3) \citep{pan2011}. 
The premise of this study is that mice with memory impairment will spend equal amounts of time in each quadrant of the maze (swimming aimlessly, so to speak), while normal mice will spend more time in the target quadrant because they can remember the location of the platform.
\begin{figure}[htbp]
\centering
\begin{tikzpicture}
    \node[circle, draw, minimum size = 4em]{Root}[sibling distance= 5 cm, level distance = 3 cm]
        child {node[circle, draw, minimum size = 4em] {$N_1$} 
        [sibling distance = 2.5 cm] 
            child{node[circle, draw, minimum size = 4em] {AQ1}
            edge from parent node[left, xshift=-0.2cm] {\Large 11.6}}
            child {node[circle, draw, minimum size = 4em] {OQ}
            edge from parent node[right, xshift=0.2cm] {\Large 10.3}}
            edge from parent node[left, xshift=-0.2cm] {\Large 8.1}
        }
        child {node[circle, draw, minimum size = 4em] {$N_2$} [sibling distance = 2.5 cm]
            child {node[circle, draw, minimum size = 4em] {AQ2}
            edge from parent node[left, xshift=-0.2cm] {\Large 5.6}}
            child{node[circle, draw, minimum size = 4em] {TQ}
            edge from parent node[right, xshift=0.2cm] {\Large 9.2}}
            edge from parent node[right, xshift=0.2cm] {\Large 11.2}
        };
    \end{tikzpicture}
 \caption{The best fitting tree with the branches labeled with the corresponding MLEs of the $\alpha$ parameters.}
 \label{fig:wmtree}
\end{figure}
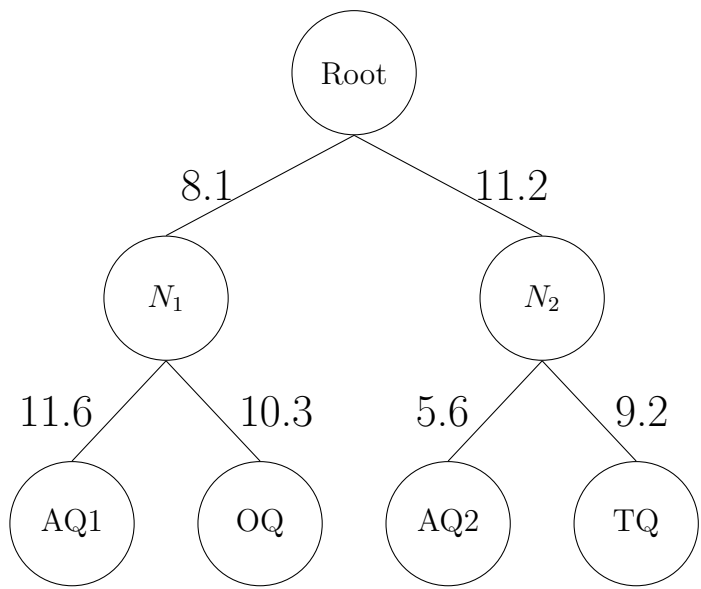
 
Figure \ref{fig:wmtree} shows the NDD fit to the water maze data using the tree-finding algorithm. The nesting structure produced in this example only slightly modifies the fit compared the fit that was determined via a brute-force search \citep{tml2025}. AQ1 and OQ variables have roughly the same means, but the variances are different, as we can see visually from the marginal fits. This is somewhat highlighted by the fact that AIC and BIC criteria choose the DD from the tree-finding algorithm. The inclusion of the nesting structure not only captures the marginal behavior well but also the correlation structure as seen below.

Table \ref{tab:fitcompare} shows the original correlation matrix juxtaposed with the correlation matrix generated from the NDD fit. Positive correlations, in bold face type, are of similar magnitude and between the same components in both matrices.
\begin{table}[ht]
\centering
\begin{tabular}{lccccccccc}
\toprule
& \multicolumn{4}{c}{Sample Correlation Matrix}& & \multicolumn{4}{c}{Correlation Matrix from NDD Fit}                       \\ 
\cmidrule(r){2-5}\cmidrule(l){7-10}
 & TQ & AQ1 & OQ & AQ2 & & TQ & AQ1 & OQ & AQ2 \\  
TQ & 1.00 & -0.67 & -0.51 & -0.32 && 1.00 & -0.54 & -0.52 & -0.29 \\ 
  AQ1 & -0.67 & 1.00 & \textbf{0.15} & -0.25 && -0.54 & 1.00 & \textbf{0.19} & -0.39 \\ 
  OQ & -0.51 & \textbf{0.15} & 1.00 & -0.27 & & -0.52 & \textbf{0.19} & 1.00 &  -0.37\\ 
  AQ2 & -0.32 & -0.25 & -0.27 & 1.00&& -0.29 & -0.39 & -0.37 & 1.00 \\ 
\bottomrule
\end{tabular}
\caption{Sample correlation matrix from the water maze data on the left compared to the correlation matrix from the NDD fit using the tree-finding algorithm. }\label{tab:fitcompare}
\end{table}

The absolute differences between the correlation matrix for the original data and the correlation matrix for the data generated from the NDD fit are similar in magnitude and sign. For example, the correlation between TQ and AQ1 for the original data is $-.67$ and the correlation between TQ and AQ1 for the generated data is $-.54$. In addition, the correlation between OQ and AQ1 is positive for both the original data and the generated data. The mean of the element-wise difference between the matrices is $.007$ and the standard deviation is $.079$, indicating that the tree finding algorithm and the exhaustive search produce similar results, even with a small sample size of $n=14$. 

\section{Discussion and Conclusion}
\label{sec:conc}
This paper introduces a data-driven framework for estimating tree structures and performing model diagnostics under the Nested Dirichlet Distribution (NDD), a flexible alternative to the Dirichlet model for compositional data. However, the NDD requires a tree structure that directly controls the key characteristics of the distribution. Our contributions address two key challenges in applying the NDD: the lack of tools for tree structure selection and the absence of diagnostic measures to assess model fit. To solve the former, we proposed a greedy tree-finding algorithm based on maximizing the likelihood. To solve the latter, we developed a novel saddlepoint approximation to the marginal distributions of the NDD, which allows the construction of pseudo-residuals and provides interpretable diagnostic plots. We further introduced a likelihood displacement measure to identify influential observations.


Our approach was validated through simulation studies and demonstrated in a behavioral application involving Morris water maze data. The methods yielded interpretable structures and identified model misfit in a way that was not possible with previous approaches. To assess the fit of an individual model, we proposed three key diagnostics that allow for investigations of the fit from both a marginal and joint perspective.  We demonstrated the overall utility of our methods with both a simulation and a real data set. For both the simulated and real data sets, the tree finding algorithm produces a reasonable tree.  By ``reasonable tree'', we mean that, in the simulation setting, the algorithm produces a tree that is either the exact tree or a slightly larger tree that generalizes the exact tree with parameter estimates that are approximately equal to those of the known tree.  In the Water Maze data setting, the algorithm produces a tree that was found using an exhaustive search where all possible trees are considered and fitted with the best model selected based on the log-likelihood \cite{tml2025}.  

The proposed saddlepoint-based diagnostics are particularly noteworthy. They extend the practical utility of the NDD by enabling residual analysis in models where traditional diagnostics are infeasible, thus offering a new pathway for model validation in complex compositional settings. By overcoming the intractability of the NDD’s marginal distributions, our approach provides applied researchers with practical tools for assessing model fit and identifying potential outliers or misfit in complex compositional data. The saddlepoint approximation achieves both high accuracy and computational efficiency, making it well suited for use in routine applied analysis. We anticipate that this innovation will encourage broader adoption of the NDD in applied compositional data settings. In addition, we have introduced a suite of freely available algorithmic tools to fit and diagnose the fit of a nested Dirichlet distribution to compositional data.  

Future work consists of methodological, computational, and software refinements and extensions. While a Likelihood ratio test has been developed to compare the population mean vectors between two NDDs, the current testing procedure assumes the tree is known in advance and that the same tree governs both populations \citep{tml2025}.  Extensive simulation studies are needed to provide insight into best practices in how a tree should be determined and how this decision affects the overall performance of the test.  We have seen in previous work that applying statistical procedures (e.g. tests for normality or homogeneity of variance) to make a decision on downstream analysis and hypothesis testing can often have an impact on type-I error control \citep{schucany2006}. Examining the overall performance of the likelihood ratio test both with and without implementing a tree finding algorithm would be of interest.  We hypothesize that selecting slightly larger trees than are necessary will correspond to a more conservative test in terms of type-I error control.  Extending the two-sample comparison methodology of \cite{tml2025} and the one-way methodology of \cite{luedeker2022} to other experimental settings, such as a factorial design, is also of interest. 

In regards to computation and software, there are instances within our methodological framework that could benefit from parallel computing to increase computational efficiency. Of particular interest will be speeding up computations with the likelihood displacement diagnostic metric as well as computing parameter estimates when working with multidimensional compositions where the number of nodes in the tree is quite large.  In the supplementary material, we have provided an R script which contains numerous functions to implement our methods to new data sets in practice.  We also provide thorough examples of their implementation using both the synthetic and Water Maze data illustrated in this manuscript.  These resources will be formally made available to the public as an R package in the future.

\bigskip
\begin{center}
{\large\bf Code and Data Availability}
\end{center}

\begin{description}

\item[GitHub repository for the tree finding algorithm:] Code to run the tree-finding algorithm and perform the diagnostic methods described in the article is found at \url{https://github.com/thestatistician88/NestedDirichlet}. The repository
also contains R Markdown files to render the results in the simulation and water maze examples. Links to the water maze data are found within the R Markdown file.

\item[Use of Generative AI:] Generative AI (ChatGPT 4o) was used to fine-tune the abstract in accordance with guidelines listed on the publisher's website for research articles.

\end{description}



\bibliographystyle{jasa3}

\bibliography{tree}
\end{document}